\documentclass[twocolumn,aps,prb,amsmath,amssymb]{revtex4} 
\usepackage{bm}
\usepackage{enumitem}
\usepackage{graphicx}
\usepackage{color,xcolor}
\usepackage{amsmath}
\usepackage{amssymb}

\newcommand{\Sec}[1]{Sec.\,\ref{#1}}

\newcommand{\ud}{{\mathrm{d}}}

\newcommand{\B}{\mbox{\tiny B}}
\newcommand{\tS}{\mbox{\tiny S}}

\newcommand{\BO}{\mbox{\tiny BO}}
\newcommand{\T}{\mbox{\tiny tot}}
\newcommand{\sol}{\mbox{\tiny sol}}
\newcommand{\SB}{\mbox{\tiny SB}}

\newcommand{\w}{\omega}
\newcommand{\ti}{\tilde}
\newcommand{\nl}{\nonumber \\}

\newcommand{\wti}{\widetilde}

\newcommand{\be}{\begin{equation}}
\newcommand{\ee}{\end{equation}}
\newcommand{\bsube}{\begin{subequations}}
\newcommand{\esube}{\end{subequations}}
\newcommand{\Eq}[1]{Eq.\,(\ref{#1})}
\newcommand{\Eqs}[1]{Eqs.\,(\ref{#1})}
\newcommand{\Fig}[1]{Fig.\,\ref{#1}}

\newcommand{\la}{\langle}
\newcommand{\ra}{\rangle}

\begin{document}

\title{Correlated vibration--solvent and Duschinsky effects on electron transfer dynamics and optical spectroscopy}

\author{Zi-Fan Zhu}
\affiliation{Hefei National Laboratory,
University of Science and Technology of China, Hefei, Anhui 230088, China}
\affiliation{Hefei National Research Center for Physical Sciences at the Microscale and Department of Chemical Physics, University of Science and Technology of China, Hefei, Anhui 230026, China}

\author{Yu Su}
\affiliation{Hefei National Laboratory,
University of Science and Technology of China, Hefei, Anhui 230088, China}
\affiliation{Hefei National Research Center for Physical Sciences at the Microscale and Department of Chemical Physics, University of Science and Technology of China, Hefei, Anhui 230026, China}

\author{Yao Wang}
\email{wy2010@ustc.edu.cn}
\affiliation{Hefei National Laboratory,
University of Science and Technology of China, Hefei, Anhui 230088, China}
\affiliation{Hefei National Research Center for Physical Sciences at the Microscale and Department of Chemical Physics, University of Science and Technology of China, Hefei, Anhui 230026, China}

\author{Rui-Xue Xu}
\email{rxxu@ustc.edu.cn}
\affiliation{Hefei National Laboratory,
University of Science and Technology of China, Hefei, Anhui 230088, China}
\affiliation{Hefei National Research Center for Physical Sciences at the Microscale and Department of Chemical Physics, University of Science and Technology of China, Hefei, Anhui 230026, China}

\author{YiJing Yan}
\affiliation{Hefei National Research Center for Physical Sciences at the Microscale and Department of Chemical Physics, University of Science and Technology of China, Hefei, Anhui 230026, China}

\date{\today}

\begin{abstract}
Understanding the effects of vibrations in electron transfer (ET) dynamics and optical spectroscopies is essential to precisely interpret the role of decoherence, especially for systems embedded in solvents. In this work, we study the correlated Duschinsky and solvent effects on ET and spectroscopy. Exploited is a novel extended dissipaton-equation-of-motion (ext-DEOM) approach, which is an exact and non-Markovian, non-perturbative method for quadratic system-bath couplings. The unified bath description, in terms of multiple Brownian oscillators (BOs), comprises the solvent modes and also intramolecular vibrations.
Both ET dynamics and spectroscopy show the complex 
interplay among linear displacements, frequency shifts, Duschinsky rotations and solvent-induced BO-mode correlations.
The reduced ET system density operator evolution is further analyzed in the context of Bloch sphere representation that is basis-set independent due to its geometric nature.
\end{abstract}

\maketitle

\section{Introduction}

Interpretations on electron transfer (ET) dynamics and optical spectroscopies 
are essential 
for exploring properties of electronic systems under the influences of vibrations and solvents.\cite{Nit06,Muk95,Lev88,Fle86,Wei21,Nit79163, Nit88489,Nit005661}
In particular, the role of vibrations, where the motions of different vibrational modes are not indeed independent but correlated via the solvent, has become increasingly recognized to be crucial for an accurate investigation on properties of molecular systems. This is especially significant for solvated systems where the solvent introduces additional complexities.\cite{Sma83515,Mor83249,Ner915433,Ole957180}

In many molecular systems, vibrational modes are strongly coupled to electronic transitions, and these couplings can be further influenced by the surrounding solvent. A prominent feature of solvated systems is the solvent influence on the decoherence dynamics and structures of spectral bands. This can be understood in terms of a dynamic solvation shell that responds/reorganizes  to the molecular motion. Additionally, the Duschinsky effect, which arises from the rotation of the normal modes upon electronic transitions/excitations, is another critical factor.\cite{Dus37551}
This effect, which describes the mixing of vibrational modes between the ground and excited states, can further complicate the analysis.\cite{Yan865908,Pen07114302,San08224311,Ma124474,Fer133597,Wan226391}

Despite its importance, an accurate, generalized and comprehensive theoretical framework that accounts for correlated vibrations and solvent effects remains elusive. Traditional method\cite{Yan865908} treats these effects separately, leading to simplifications which may not fully capture the complexity of real molecular systems. To address this, we present a combined approach that integrates the solvent and Duschinsky effects within a unified framework, using the extended dissipaton-equation-of-motion (ext-DEOM) formalism with quadratic environment  (bath) couplings.\cite{Xu17395,Xu18114103} In our previous work,\cite{Xu18114103} only changes of frequencies of vibrational normal modes upon the electronic excitation are considered. This paper generalizes to include the Duschinsky rotations, together with the correlated solvent effects. 

The DEOM formalism has been established as an exact and nonperturbative approach for open quantum systems.\cite{Yan14054105,Wan22170901}It adopts dissipaton as quasi-particle to characterize the statistical property of thermal bath. DEOM recovers the hierarchical equations of motion (HEOM) method \cite{Tan906676,Yan04216,Xu05041103,Jin08234703,Tan20020901,Zha24e1727} for the reduced system dynamics.
HEOM was constructed via the derivative on the Feymann--Vernon influence functional path integral,\cite{Fey63118,Tan906676,Xu05041103} for bath couplings satisfying Gaussian statistics modeled as harmonic oscillators  linearly coupled to the system. 
The dissipaton theory is more convenient and straightforward for bath collective dynamics and polarizations as well as further extension to nonlinear bath couplings which is of non-Gaussian statistics.\cite{Xu17395, Xu18114103,Su23024113,Su254107} 

In this work, we apply the ext-DEOM to a system with two electronic states, focusing on the combined effects of correlated vibrations and solvent interactions, and the Duschinsky rotations. We show how these factors affect the ET decoherence dynamics and optical spectra. 
The environment is composed of intramolecular vibration modes and the solvent.
To apply the ext-DEOM, one key step is to obtain the bath coupling
descriptors with the linear displacement ansatz\cite{Xu18114103}
that is generalized in this paper to include the Duschinsky rotation between multimodes. 
We elaborate how an accurate description can be achieved, thus offering an exact approach with the ext-DEOM to interpret both ET process and absorption spectra. Furthermore the ET decoherence dynamics is visualized
via the Bloch sphere representation that is basis-set independent due to its geometric nature. Our results highlight the importance of a comprehensive treatment of correlatied vibrational and solvent effects for reliable predictions in complex systems.

The remainder of this paper is organized as follows.
We present the total composed Hamiltonian in \Sec{thsec2}, and the derivation to obtain the bath coupling
descriptors with the linear displacement ansatz in \Sec{thdesc}.
In \Sec{thdeom}, we present the ext-DEOM formalism that will be adopted in the simulation.
Numerical demontrations are presented in \Sec{thsec3}.
Finally, we summarize the paper in \Sec{thsec4}.
Throughout this paper, we set the Planck constant and Boltzmann constant as units ($\hbar=1$ and $k_B=1$), and $\beta=1/T$, with $T$ being the temperature.

\section{Theory}
\label{sectheo}

In this section, we start from the Hamiltonian of total composition
where the intramolecular vibrational modes involve the Duschinsky rotation upon the electronic state transition. After that we derive the linear and quadratic bath coupling descriptors via the linear displacement ansatz. The extended dissipaton formalism applied to this system is then presented.

\subsection{Total Hamiltonian}
\label{thsec2}

Consider a molecular system, consisting of two electronic states and vibrational modes $\{\hat q_n\}$ and $\{\hat q'_n\}$ on the ground ($|g\ra$) and the excited ($|e\ra$) surfaces, respectively, embedded in solvent environments. The total molecular composite  Hamiltonian reads
\begin{align}\label{Htot}
 H_{\T} &= H_g|g\ra\la g| + (H_e+\w_{eg}) |e\ra\la e|+\hat V,
\end{align}
where $\hat V\equiv V(|e\ra \la g|+|g\ra\la e|)$ is the interstate coupling that is assumed independent of nuclear modes.
The nuclear Hamiltonians are modelled by Brownian oscillators (BOs),
\bsube \label{wy_0115} 
\begin{align}\label{wy_0115_a}
H_g&=\sum_{n=1}^{N}
\frac{\Omega_{n}}{2}\big(\hat p_n^2+\hat q_n^2\big)
\nl & \quad
+ \!\sum_{k=1}^{N_{\sol}}\!\frac{ \w_k}{2}\!
\bigg[   \ti p^2_k+\!
\Big( \ti  x_k-
\!
\sum_n
\frac{ c_{nk}}{\w_k}\hat q_n\Big)^2
\bigg]
,
\\ 
H_e&=\sum_{n=1}^{N}
\frac{\Omega'_{n}}{2}\big({\hat p}_n^{\prime 2}+\hat q_n^{\prime 2}\big)
\nl &\quad
+ \!\sum_{k=1}^{N_{\sol}}\!\frac{ \w_k}{2}\!
\bigg\{  \ti  p^2_k+\!
\Big[ (\ti  x_k-\ti d_k)
-\sum_n
\frac{ c'_{nk}}{\w_k}\hat q'_n\Big]^2
\bigg\},\label{wy_0115_b}
\end{align}
\esube
with $\{\ti d_k\}$ being the linear displacements of the solvent modes $\{\ti x_k\}$.
As a continuous solvent model, the solvent modes 
are assumed irrotational and frequencies 
 $\{\w_k\}$ unchanged.

The molecular vibrational modes in two surfaces involve the displacements ($\{d_m\}$), frequency shifts  ($\{\Omega'_n\}$ versus $\{\Omega_m\}$), and Duschinsky rotation ($\{\bar S_{nm}\}$), related by
\begin{align}\label{coorditrans1}
\hat q_n'=&\sum_{m} \bar S_{nm}(\Omega'_n/\Omega_{m})^{\frac{1}{2}}( \hat q_{m}- d_{m}).
\end{align}
In matrix  form it reads
 \be \label{qprime}
\hat {\bm q}'=
{{\bm \Omega}'}^{\frac{1}{2}}
\bar{\bm S}
{\bm \Omega}^{-\frac{1}{2}}
(\hat {\bm q}-{\bm d})\equiv {\bm S}(\hat {\bm q}-{\bm d}),
 \ee
with $ 
\bar {\bm S}^{T}\bar {\bm S}={\bm I}
$. Here
\begin{align}\label{rotat}
{\bm S}
\equiv{{\bm \Omega}'}^{\frac{1}{2}}
\bar{\bm S}
{\bm \Omega}^{-\frac{1}{2}},
\,\,
\end{align}
${\bm \Omega}
={\rm diag}\{\Omega_1
,\cdots,\Omega_N
\}$ and ${\bm \Omega}'
={\rm diag}\{\Omega^{\prime}_1
,\cdots,\Omega^{\prime}_N
\}$. 
For later use, we denote also $
{\bm S}'\equiv  {\bm\Omega}^{\frac{1}{2}}
\bar{\bm S}^T
{{\bm \Omega}'}^{-\frac{1}{2}}
={\bm S}^{-1}$.

In \Eq{wy_0115}, we further assume the relation between the coupling coefficients 
\begin{align}\label{para}
c'_{nk}
=\sum_{m}
S^{\prime T}_{nm}
c_{mk},
\end{align}
which results in
\begin{align}\label{solvibcoup}
\sum_{nk}c'_{nk}\hat{q}'_n (\ti x_{k}-\ti d_k)
=
\sum_{nk}
c_{nk}
(\hat q_n - d_n)
(
\ti x_{k} -\ti d_k
)
.
\end{align}
Physically this amounts to that the change of overall vibration--solvent interaction depends just on the displacements upon the electronic transition. The physical consideration here is based on that the solvent is a continuous fluid with infinite number of degrees of freedom. Finally we note in case that the Duschinsky effect of certain neighboring shell of solvent is not negligible, it may be treated in the way as the intramolecular modes.

To complete the BO description, we shall also characterize the response functions, 
\begin{align}
\label{chig}
{\bm\chi}_g(t)
\equiv
\{
\chi_{g,mn}(t)
\equiv
i
\la[\hat{q}_m(t),
\hat{q}_n(0)
]\ra_{g}
\},
\end{align}
Here, $ \langle \hat O \rangle_g \equiv {\rm tr}_{g}(\hat{O}
e^{-\beta H_{g}}
) /{\rm tr}_g(
e^{-\beta H_{g}}
) $ and 
$
\hat q_m(t)\equiv e^{iH_{g}t}
\hat q_m
e^{-iH_{g}t}
$.
In terms of frequency resolution,
$\ti f(\w)\equiv\int_{0}^{\infty}\!\!\ud t \,e^{i\w t}\!f(t)$,
\Eq{chig} is resolved as the BO form,\cite{Wei21,Yan05187,Ton23024117}
\begin{align}\label{chi_w}
{\wti{\bm\chi}}_g(\w)
&=\Big[{\bm\Omega}^2-\w^2-i{\omega}
{\wti{\bm\zeta}}(\w)
\Big]^{-1}{\bm\Omega},
\end{align}
where $\wti{\bm\zeta}(\w)$ is the solvent frictional resolution.\cite{Wei21,Yan05187}
For later use, we define
\bsube
\be \wti\chi^{(+)}_{g,mn}(\w)\equiv [\wti\chi_{g,mn}(\w)+\wti\chi^{\ast}_{g,nm}(\w)]/2
\ee
and
\be \label{10b_0416}\wti\chi^{(-)}_{g,mn}(\w)\equiv [\wti\chi_{g,mn}(\w)-\wti\chi^{\ast}_{g,nm}(\w)]/(2i).
\ee
\esube
They are related via the Kramers-Kronig relation \cite{Wei21,Yan05187}
\bsube
\be \label{11a_0416}
\wti\chi^{(+)}_{g,mn}(\w)=-\frac{1}{\pi}{\cal P}\int_{-\infty}^{\infty}\!\!{\rm d}\w'\,\frac{\wti\chi^{(-)}_{g,mn}(\w')}{\w-\w'}
\ee
and
\be 
\wti\chi^{(-)}_{g,mn}(\w)=\frac{1}{\pi}{\cal P}\int_{-\infty}^{\infty}\!\!{\rm d}\w'\,\frac{\wti\chi^{(+)}_{g,mn}(\w')}{\w-\w'}
\ee
\esube
where ${\cal P}$ denotes the principle integration.

\subsection{Bath coupling
descriptors}
\label{thdesc}

To apply methods concerning open systems, we need to recast the total Hamiltonian into the system-plus-environment (bath) form, $H_{\T}=H_{\tS}+H_{\tS\B}+h_{\B}$. 
Assuming the electronic system initially at the ground state $|g\ra$ equilibrated 
with its surroundings $H_g$, we rewrite \Eq{Htot} as
\begin{align}\label{wy_0421_12}
 H_{\T}
=\w_{eg}|e\ra\la e|+\hat V+(H_{e}-H_{g})|e\ra\la e|+H_{g}.
\end{align}
To continue, let us denote 
\begin{align}
X_{n} =\sum_k c_{nk}\ti x_k\ \ \text{and}\ \ 
D_{n} = \sum_{k}c_{nk}\ti d_k.
\end{align}
Inferred from \Eqs{wy_0115_a} and (\ref{wy_0115_b}),
the renormalized frequencies of BOs according to the two states are respectively 
\begin{align}
\label{eta}
{\ti\eta}_{mn}\equiv\sum_k\frac{ c_{mk} c_{nk}}{\w_{k}}\ \  \text{and}  \ \  
{\ti\eta}'_{mn}\equiv\sum_k\frac{c'_{mk} c'_{nk}}{\w_{k}}.
\end{align}
In terms of the matrix form $\ti {\bm \eta}$ and $\ti {\bm \eta}'$, they are related by $
{\ti{\bm\eta}}' = {\bm S}^{\prime T}
{\ti{\bm\eta}}
{\bm S}'$. 
We thus obtain
\begin{align}\label{diff_0416}
H_e-H_g&=
\frac{1}{2}
\hat {\bm q}^T
(
{\bm S}^T
{\bm\Omega'}
{\bm S}-
{\bm\Omega}
)
\hat{\bm q}
-
\hat{\bm q}^T
{\bm S}^T
{\bm\Omega'}
{\bm S}{\bm d}
\nl
&\quad-
\hat{\bm q}^T
\ti{\bm\eta}
{\bm d}+ 
\hat{\bm q}^T
{\bm D}+
\frac{1}{2}
{\bm d}^T{\bm S}^T
{\bm\Omega'}
{\bm S}{\bm d}
+
\frac{1}{2}
{\bm d}^T
\ti{\bm\eta}
{\bm d}
\nl &\quad
+
{\bm d}^T{\bm X}
-
\ti {\bm x}^T{\bm\w}{\ti{\bm d}}-
{\bm d}^T
{\bm D}
+
\frac{1}{2}
{\ti{\bm d}}^T{\bm\w}{\ti{\bm d}}.
\end{align}

To derive the linear and quadratic bath coupling descriptors, we adopt in the following the linear displacement ansatz proposed in Ref.\,\onlinecite{Xu18114103} for the case of single solvation bath mode. This ansatz arises from considering the limit that the quadratic coupling term vanishes. In the present work that is $\bm {\bar S}=1$ and ${\bm \Omega}'={\bm \Omega}$, leading to $\hat {\bm q}'=\hat{\bm q}-{\bm d}$.
The mutually coupled harmonic oscillators in \Eq{wy_0115} can be transformed into independent normal modes, $\{{\rm x}_j\}$, and displaced as $\{{\rm x}_j-\bar{\rm d}_j\}$ at the $|e\ra$ state. We have
\begin{align}\label{16_0416}
(H_e-H_g)_{\rm no\  quadraitc} =
\sum_{j=1}^{N+N_{\sol}}\!\!\Big(\frac{1}{2}\bar{\w}_j
{\bar{\rm d}}^2_j-\bar{\w}_j
\bar{\rm d}_j{\rm x}_j\Big).
\end{align}
We can always define (mathematically  a problem of more unknowns than equations/restrictions)
\begin{align}\label{q_term}
\sum_j\bar{\w}_j
\bar{\rm d}_j{\rm x}_j
\equiv\sum_{n=1}^{N}(2\lambda_n\Omega_n)^{\frac{1}{2}}
\sum_j\bar{\rm c}_{nj}{\rm x}_j
=\sum_{n=1}^{N}(2\lambda_n\Omega_n)^{\frac{1}{2}}\hat{\rm q}_n,
\end{align}
where
\begin{align}\label{d_term}
\bar{\rm d}_j
=
\frac{1}{\bar{\w}_j}\sum_{n=1}^N(2\lambda_n\Omega_n)^{\frac{1}{2}}\bar{\rm c}_{nj},
\end{align}
with $\lambda_n$ to be determined later [cf.\,\Eq{24_0416}].
At the microscopic level, the spectral density, the counterpart of \Eq{10b_0416}, reads 
\begin{align}\label{19_0416}
\wti\chi^{(-)}_{g,mn}(\w) =
\frac{\pi}{2}\sum_{j}{\rm\bar c}_{mj}{\rm\bar c}_{nj}\Big[\delta(\w-\bar{\w}_j)-\delta(\w+\bar{\w}_j)\Big].
\end{align}
Note that $\wti\chi^{(-)}_{g,mn}(\w=0)=0$. Therefore, due to \Eq{11a_0416}, we obtain
\be \label{20_0416}
\wti\chi^{}_{g,mn}(0)=\wti\chi^{(+)}_{g,mn}(0)=\frac{1}{\pi}{\cal P}\int_{-\infty}^{\infty}\!\!{\rm d}\w\,\frac{\wti\chi^{(-)}_{g,mn}(\w)}{\w}.
\ee
Equations (\ref{19_0416}) and (\ref{20_0416}) together give
\be 
\wti\chi^{}_{g,mn}(0)=\sum_{j}\frac{{\rm\bar c}_{mj}{\rm\bar c}_{nj}}{\bar\w_j}.
\ee
On the other hand, from \Eq{chi_w} we have 
$
\wti{\bm \chi}_{g}(0)={\bm \Omega}^{-1}
$. 
As a result,
\begin{align}\label{22_0416}
\frac{1}{2}\sum_j\bar{\w}_j
\bar{\rm d}^2_j
=&
\frac{1}{2}\sum_j\frac{1}{\bar{\w}_j}\sum_{mn}
(2\lambda_n\Omega_n)^{\frac{1}{2}}(2\lambda_m\Omega_m)^{\frac{1}{2}}\bar{\rm c}_{mj}\bar{\rm c}_{nj}
\nl=&
\sum_{n}\lambda_n.
\end{align}
By \Eqs{q_term} and (\ref{22_0416}), \Eq{16_0416} is thus expressed as
\be \label{23_0416}
(H_e-H_g)_{\rm no\  quadratic}=\sum_{n=1}^{N}\big[\lambda_n-(2\lambda_n\Omega_n)^{\frac{1}{2}}\hat{\rm q}_n\big].
\ee
By the same way of 
linear combination as $\hat{\rm q}_n=\sum_j\bar{\rm c}_{nj}{\rm x}_j
$, we have [cf.\,\Eq{d_term}]
\begin{align}\label{24_0416}
{\rm d}_n = \sum_{j}\bar{\rm c}_{nj}\bar{\rm d}_j
= \sum_{j}
\frac{\bar{\rm c}_{nj}}{\bar{\w}_j}\sum_m(2\lambda_m\Omega_m)^{\frac{1}{2}}\bar{\rm c}_{mj}
=\Big( \frac{2\lambda_n}{\Omega_n} \Big)^{\frac{1}{2}}.
\end{align}
With this ansatz elaborated in the above, comparing \Eq{23_0416} to  \Eq{diff_0416} at the scenario without quadratic couplings ($\bm {\bar S}=1$ and ${\bm \Omega}'={\bm \Omega}$), 
we obtain
\begin{align}\label{LDA}
\begin{split}
& {\bm d}^T
 \ti{\bm\eta}
 {\bm d}
 -2{\bm d}^T{\bm D}
 +\ti{\bm d}^T
{\bm{\omega}}
\ti{\bm d}=0, 
\\ & 
 {\bm q}^{T}\ti{\bm\eta}{\bm d}-{\bm q}^T{\bm D}-{\bm d}^T{\bm X}+{\bm{x}}^T
{\bm{\omega}}{\ti{\bm d}}=0.
\end{split}
\end{align}
This finishes the multi-mode generalization involving Duschinsky rotation of the linear displacement ansatz brought forward for the single-mode scenario in Ref.\,\onlinecite{Xu18114103}.
The derivation in Ref.\,\onlinecite{Xu18114103} adopts further a linear displacement mapping to obtain the descriptors. The mapping is not needed for the modeled system in this work.
Note that simulations based on the full form of \Eq{diff_0416} without adopting the linear displacement ansatz can also be carried out.
The treatments on the entangled vibration--solvent terms will be similar as those in Ref.\,\onlinecite{Che21244105}.

By using \Eq{LDA}, \Eq{diff_0416} then readily 
leads to 
\begin{align}\label{hatU_def}
  H_e-H_g
=\alpha_0+{\bm\alpha}_1 \cdot \hat{\bm q}+\hat{\bm q}^{T}{\bm\alpha}_2\hat{\bm q},
\end{align}
with the bath coupling descriptors \bsube\label{alpha}
\begin{align}
&\alpha_0=\frac{1}{2}{\bm d}^{T}
{\bm S}^{T}
{{\bm \Omega}'}
{\bm S}
{\bm d}, 
\\ 
&{\bm \alpha}_1
=-
{\bm S}^{T}
{{\bm\Omega}'}
{\bm S}
{\bm d},
\\
&{\bm \alpha}_2=\frac{1}{2}
\big(
{\bm S}^{T}
{{\bm\Omega}'}
{\bm S}
-
{{\bm\Omega}}
\big).
\end{align}
\esube
Here we assume the electronic system is initially at the ground state $|g\ra$ equilibrated 
with its surroundings $H_g$. 
For the other condition that the system is initially equilibrated at $|e\ra$ with $H_e$,
there is correspondingly
\be\label{emission}
 H_{\T}
=\w_{eg}|e\ra\la e|+\hat V+(H_{g}-H_{e})|g\ra\la g|+ H_{e},
\ee
where
\be\label{h_diff_em}
H_g-H_e=\alpha'_0+{\bm\alpha}_1^{\prime }\cdot \hat{\bm q}'+\hat{\bm q}^{\prime T}{\bm\alpha}'_2\hat{\bm q}^{\prime},
\ee
with bath coupling descriptors reading
\bsube\label{alphap}
\begin{align}
&\alpha_0'=\frac{1}{2}
{\bm d}^{ T}
{{\bm\Omega}}
{\bm d},
\\
&{\bm \alpha}'_1=
{\bm S}^{\prime T}
{{\bm \Omega}}{\bm d},
\\
&{\bm \alpha}'_2=\frac{1}{2}
\big(
{\bm S}^{\prime T}
{{\bm\Omega}}
{\bm S}^{\prime}-{{\bm\Omega}'}
\big).
\end{align}
\esube
For single-mode scenarios, \Eqs{alpha} and (\ref{alphap}) reduce to the results in Ref.\,\onlinecite{Xu18114103}.

\subsection{Dissipaton theory with linear and quadratic environment couplings}\label{thdeom}

The quantum dissipative dynamics method starts from the total system--plus--bath  composite Hamiltonian being of the form
$
H_{\T}=H_{\tS}+H_{\tS\B}+h_{\B}$.
Concerning \Eqs{wy_0421_12} and (\ref{hatU_def}),
$H_{\tS}=(\w_{eg}+\alpha_0)|e\ra\la e|+\hat V$ and
\begin{align}\label{0421_31}
H_{\SB}=
 \hat Q \Big(
{\bm\alpha}_1 \cdot \hat{\bm q}+\hat{\bm q}^{T}{\bm\alpha}_2\hat{\bm q}
\Big),
\end{align}
where
$\hat{Q}=
|e\ra\la e|$.
The harmonic bath $h_{\B} = H_g$
constitutes the Gauss--Wick's environment ansatz
where the environmental influence is fully characterized by 
the correlation functions, 
$\{\la\hat{q}_{m}(t)
\hat{q}_{n}(0)\ra_{\B}=\la\hat{q}_{m}(t)
\hat{q}_{n}(0)\ra_{g}\}$.
They are related to the spectral densities $\{{\chi}^{(-)}_{g;mn}(\w) \}$ via the fluctuation--dissipation theorem, \cite{Wei21}
\begin{align}\label{FDT}
\la\hat{q}_{m}(t)
\hat{q}_{n}(0)\ra_{\B}
&=\frac{1}{\pi }
\!\int^{\infty}_{-\infty}\!\!{\rm d}\omega\,
\frac{e^{-i\omega t}
{\chi}^{(-)}_{g;mn}(\w)
}
{1-e^{-\beta\omega}}\nl
&\simeq\sum^K_{k=1}\eta_{mn k} e^{-\gamma_{k} t}
,
\end{align}
where
\begin{align}\label{Jw}
{\chi}^{(-)}_{g;mn}(\w) 
\equiv
\frac{1}{2}
\int_{-\infty}^{\infty}
\!\!
\ud t\, e^{i\w t} 
\la [\hat{q}_m(t),
\hat{q}_n(0)
]\ra_{g}.
\end{align}
The exponential series expansion
of \Eq{FDT} can be achieved by using the time--domain Prony fitting decomposition
scheme.\cite{Che22221102}  
The time-reversal relation of correlation functions is given by
\begin{align} \label{FBt_corr}
\la\hat{q}^{\B}_{n}(0)\hat{q}^{\B}_{m}(t)\ra_{\B}
=
\la\hat{q}_{m}(t)
\hat{q}_{n}(0)\ra^{\ast}_{\B}
=
\sum^{K}_{k=1}\eta_{mn\bar k}^{\ast} e^{-\gamma_{k} t}.
\end{align}
The exponents $\{\gamma_k\}$ in \Eqs{FDT} and (\ref{FBt_corr})  
must be either real or complex conjugate paired, and thus
one may determine $\bar k$ in the index set $ \{k=1,2,...,K\}$
by the pairwise equality $\gamma_{\bar k}=\gamma_{k}^{\ast}$.
It is the exponential series expansion in \Eqs{FDT} and (\ref{FBt_corr}) that inspired the idea of relating each exponential mode of correlation function to a statistical  quasi-particle, i.e., a \emph{dissipaton}.\cite{Yan14054105,Wan22170901}

The dissipaton theory begins with
the \emph{dissipatons decomposition} that reproduces the correlation function in \Eqs{FDT} and (\ref{FBt_corr}).  
It decomposes $\hat{q}_m$ into a number of dissipaton operators, $\{\hat{f}_{m k} \}$, as
\begin{align}\label{hatFB_in_f}
\hat q_{m}=\sum^K_{k=1}  \hat{f}_{m k},
\end{align}
reproducing \Eq{FDT} and (\ref{FBt_corr}) by setting
\bsube\label{eq9}
\begin{align}\label{fx_corr}
\la \hat{f}^{\B}_{m k}(t)\hat{f}^{\B}_{n j}(0)\ra_{\B}=\delta_{k j}\eta_{mn k} e^{-\gamma_{k}t},
\\ \label{fx_corr_nu_rev}
\la \hat{f}^{\B}_{n j}(0)\hat{f}^{\B}_{m k}(t)\ra_{\B}=\delta_{k j} \eta_{mn\bar k}^{\ast} e^{-\gamma_{k}t},
\end{align}
\esube
with $\hat{f}^{\B}_{n k}(t)\equiv e^{ih_{\B}t}\hat{f}_{n k}e^{-ih_{\B}t}$.
Each forward--backward pair of dissipaton correlation functions
is specified by a single--exponent $\gamma_k$. 
In accordance with the dissipatons decomposition,  the dynamical variables in DEOM are
the dissipaton density operators
(DDOs), 
\be \label{DDO}
  \rho^{(n)}_{\bf n}(t)
\equiv {\rm tr}_{\B}\Big[
  \big(\prod_{m k}\hat{f}_{m k}^{n_{m k}}\big)^{\circ}\rho_{\T}(t)
 \Big].
\ee
Here, $n=\sum_{m k}n_{m k}$, with $n_{m k}\geq 0$
for the bosonic dissipatons.
The product of dissipaton operators inside $(\cdots)^\circ$
is \emph{irreducible},  which satisfies
$(\hat{f}_{m k}\hat{f}_{n j})^{\circ}
=(\hat{f}_{n j}\hat{f}_{m k})^{\circ}$
for bosonic dissipatons.
Each $n$--particles DDO, $\rho^{(n)}_{\bf n}(t)$, is associated with
an ordered set of indexes, ${\bf n}\equiv \{n_{m k}\}$.
Denote for later use  ${\bf n}^{\pm}_{m k}$ and ${\bf n}^{\pm,\pm}_{m k,m'k'}$ which differ from ${\bf n}$ only
at the specified dissipatons with their occupation numbers
$\pm 1$.
The reduced system density operator is the zeroth-tier DDO, 
$\rho_{\bf 0}^{(0)}(t)=\rho_{0\cdots 0}^{(0)}(t)= \rho_{\tS}(t)$.

The equation of motion for DDOs 
including both linear and quadratic bath couplings, i.e., the ext-DEOM, is obtained as\cite{Xu17395,Xu18114103}
\begin{align} \label{gen_DEOM}
\dot\rho^{(n)}_{\bf n} \!
= & - \Big[i{\cal L}_{\tS} + \gamma_{\bf n} + i
\Big(
\alpha_0+
\sum_{mm'}
\alpha_{2mm'}
\la \hat q_{m }\hat q_{m'}
\ra_{\B}
\Big)
{\cal A} \Big]
\rho^{(n)}_{\bf n} 
\nl &
-i\sum_{m k} 
\alpha_{1m}
\Big[
{\cal A}
\rho^{(n+1)}_{{\bf n}^{+}_{m k}}
+
\sum_{m'}
n_{m'k}
{\cal C}_{mm'k}
\rho^{(n-1)}_{{\bf n}^{-}_{m'k}}
\Big]
\nl & 
-i \!\!
\sum_{mm'kk'}\Big[
n_{mk}{\cal B}_{mk,m'}
 \rho^{(n)}_{{\bf n}^{-,+}_{mk, m'k'}}
+\alpha_{2mm'}
{\cal A}
\rho^{(n+2)}_{{\bf n}^{+,+}_{m k,m'k'}}
\nl &
 +
\alpha_{2mm'}n_{m k} 
(n_{m' k'}\!-\!\delta_{mm'}
\delta_{kk'})
{\cal D}_{mm' kk'}
\rho^{(n-2)}_{{\bf n}^{-,-}_{mk,m' k'}}\Big].
\end{align}
Here, $\gamma_{\bf n}\equiv \sum_{m k}n_{m k}\gamma_k$, ${\cal L}_{\tS} \hat O \equiv 
[H_{\tS}, \hat O]$, ${\cal A} \hat O \equiv [\hat Q, \hat O]$, and other involved superoperators are defined as
\begin{align} \label{def_ABC}
 &{\cal B}_{mk,m'}\hat O \equiv 2\sum_{m''}\alpha_{2m''m'}{\cal C}_{m''m k}\hat O,
\nl
&{\cal C}_{mm' k}\hat O \equiv \eta_{mm'k} \hat Q\hat O
- \eta^{\ast}_{mm'\bar k} \hat O \hat Q,
\nl
&{\cal D}_{mm'kk'} \hat O \equiv 
\sum_{ll'}\big(\eta_{mlk}
\eta_{m'l'k'}
\hat Q \hat O 
-
\eta^{\ast}_{ml \bar{k}}
\eta^{\ast}_{m'l'\bar{k'}}
\hat O 
\hat Q\big).
\nonumber
\end{align}
In \Eq{gen_DEOM}, the DDOs form a hierarchical structure with couplings not only between adjacent layers $(n\pm1)$ but also between next-nearest-neighbor layers $(n\pm 2)$.
The latter arises from the quadratic coupling term in the system-environment interaction [cf.\,$\alpha_2$-term in \Eq{0421_31}]. Equation (\ref{gen_DEOM}) is obtained via the dissipaton algebra, and its exactness has also been verified through comparisons with the stochastic-field method \cite{Che21174111} and the core-system approach.\cite{Che23074102} Notably, the ext-DEOM can be applied not only to compute real-time dynamics but also to study imaginary-time evolution and non-equilibrium thermodynamic properties. \cite{Che23074102}

\begin{figure}[t!]
\centering
\includegraphics
[width=0.45
\textwidth]
{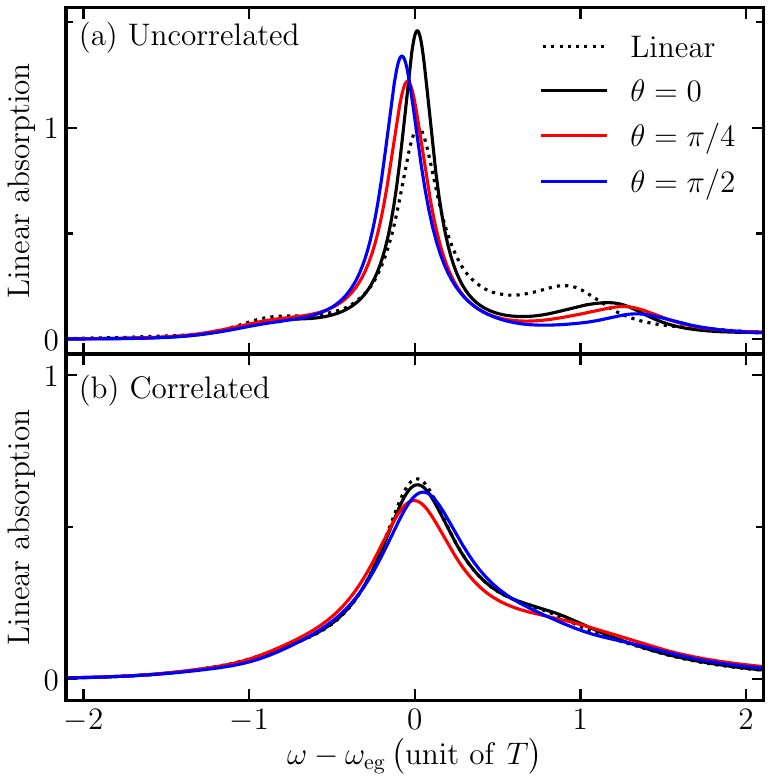}
\caption{
The evaluated absorption spectra with two low-frequency modes.
All values are uniformly scaled by the peak amplitude of linear case in (a)-panel.}
\label{fig1}
\end{figure}

\section{Numerical demonstration}
\label{thsec3}

In the following demonstrations, 
we set the temperature $T$ as the unit.
We consider two vibration modes under Duschinsky transformation characterized by
\begin{equation}
\label{rotation_m}
 \bar{\bm S} =  
\begin{pmatrix}
\cos\theta\  &
-\sin\theta \\
\sin\theta\ &\cos\theta 
\end{pmatrix}.
\end{equation}
We select the angle to be $\theta=0$, $\pi/4$, $\pi/2$.
The dimensionless displacements $d_1$ and $d_2$ in \Eq{coorditrans1} are both selected to be $-0.5$.
For the solvent friction influence, we assume the
white-noise limit for $\wti{\bm \zeta}(\w)$ in \Eq{chi_w}, i.e., $\wti{\bm \zeta}(\w)\approx \wti{\bm \zeta}(\w=0)
=\int_0^{\infty}\!\!\ud t\, {\bm\zeta}(t)
\equiv {\bm \Gamma}$.
Note that in \Eq{eta}, $\wti{\bm\eta} = {\bm\Omega}^{-1}{\bm\zeta}(t=0)$.
We will choose
$\Gamma_{11}=\Gamma_{22} = 0.3\,\Omega_1$ ($\Omega_1$ will be specified later), corresponding to underdamped cases, while 
$\Gamma_{12}=\Gamma_{21} = \Gamma_{11}/2$ and 0 represent correlated and uncorrelated 
vibration--solvent cases, respectively.

\begin{figure}[t!]
\centering
\includegraphics
[width=0.45
\textwidth]
{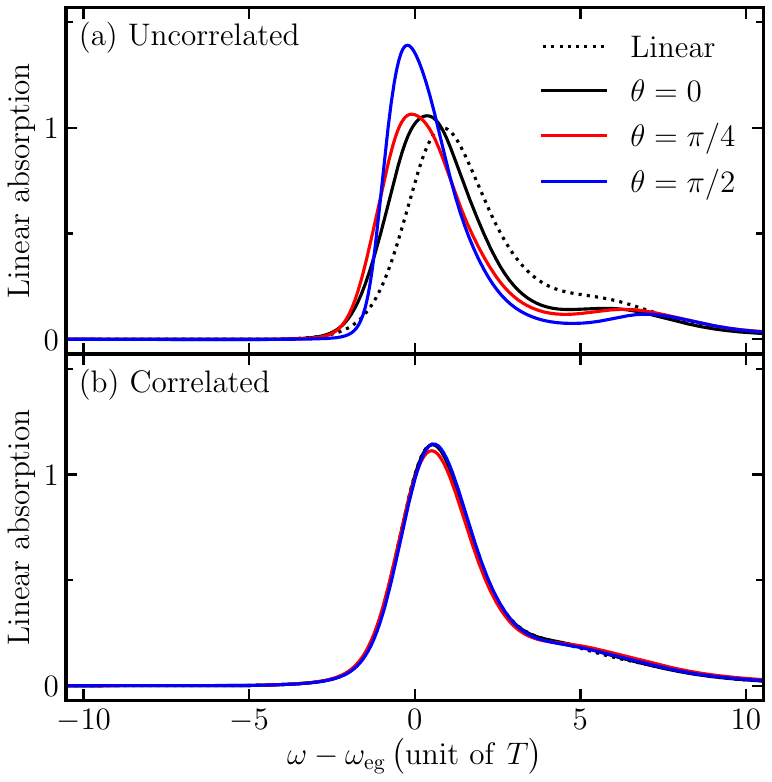}
\caption{
The evaluated absorption spectra with two high-frequency modes and one additional overdamped BO mode.
All values are uniformly scaled by the peak amplitude of linear case in (a)-panel.}
\label{fig2}
\end{figure}

\subsection{Linear absorption spectroscopy}\label{num:a}

In this subsection, we focus on the absorption spectra and set $\hat V = 0$ in \Eq{Htot}. 
Figure \ref{fig1} depicts the evaluated absorption spectra with two underdamped BO modes, $(\Omega_1,\Omega_2)/T=(0.97,1.16)$ and $(\Omega'_1,\Omega'_2)/T=(1.02,1.11)$ in the ground and excited states, respectively. 
Included for comparison is also the linear coupling counterpart, ($\bm {\bar S}=1$ and ${\bm \Omega}'={\bm \Omega}$).
In both \Fig{fig1}(a) and \Fig{fig1}(b) panels, we can observe one main peak of the electronic state transition and secondary peak with additional vibrational state excitation. 
The positions of these peaks vary with Duschinsky rotation angle $\theta$.
The spectra are
broadened for $\Gamma_{12}\neq0$; seen from \Fig{fig1}(b) with respect to \Fig{fig1}(a).
This indicates stronger decoherence induced by solvent induced BO modes correlation, which will be also observed in the ET dynamics (cf.\,\Sec{numB}).

Figure \ref{fig2} depicts the evaluated absorption spectra  with two underdamped BO modes and one additional overdamped BO mode which is neither rotated nor bath-induced correlated with the other two modes.
The two underdamped BO modes are of $(\Omega_1,
\Omega_2
)/T=(4.83,
5.80
)$ in the ground state and $(\Omega'_1, \Omega'_2)/T=(5.08,5.55)$ in the excited state, with displacements $d_1=d_2=-0.5$.
The overdamped mode that introduces additional broadening is of
$\Gamma_3/(2\Omega_3)=5$, with 
$(\Omega_3, \Omega'_3)/T=(2.90,3.15)$ and
$d_3=-0.5$.
The peaks in \Fig{fig2}(a) (uncorrelated scenario) still exhibit observable Duschinsky effect, with varied values of $\theta$, in comparison to \Fig{fig2}(b) (correlated scenario).

\begin{figure}[!t]
\centering
\includegraphics
[width=0.45
\textwidth]
{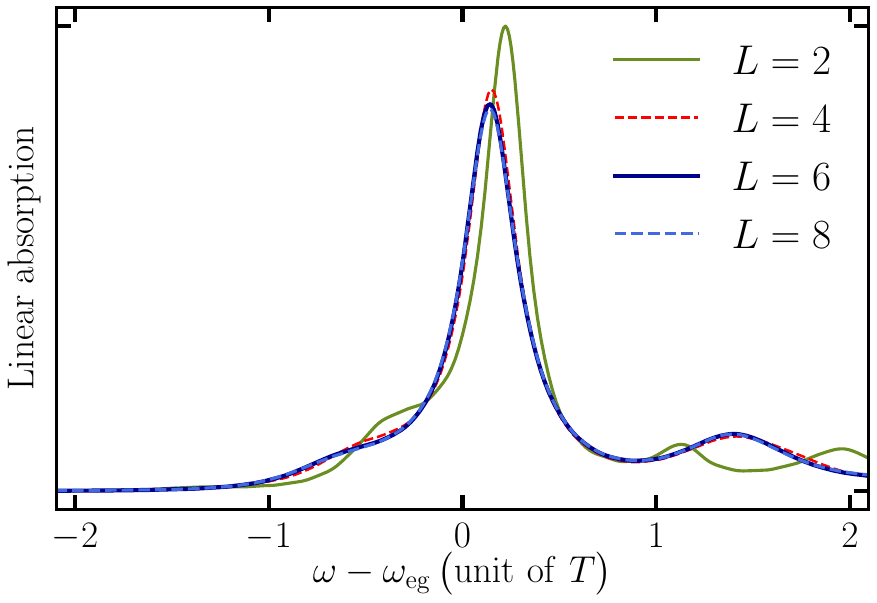}
\caption{
The evaluated absorption spectra with different truncation tier $L$. The parameters are same as those in \Fig{fig1} with $\theta=\pi/3$.}
\label{fig3}
\end{figure}

In \Fig{fig3} we exhibit the absorption spectra, exemplified with $\theta=\pi/3$, calculated at different truncation tiers $L$ by setting $\rho^{(n>L)}=0$ in \Eq{gen_DEOM}. Other parameters are same as those in \Fig{fig1}. 
The result of $L=2$ corresponds to that of the quantum master equation involving quadratic bath couplings. 
As an exact and nonperturbative theory, DEOM converges rather rapidly and monotonically with respect to truncation level.\cite{Li12266403}
Higher--frequency--shift components converge earlier.
The $L=4$ would be enough except for the zero--frequency--shift peak, while the result of $L=6$ is numerically exact in the present study.
All reported numerical results in this paper are converged.

\begin{figure}[!t]
\centering
\includegraphics
[width=0.45
\textwidth]
{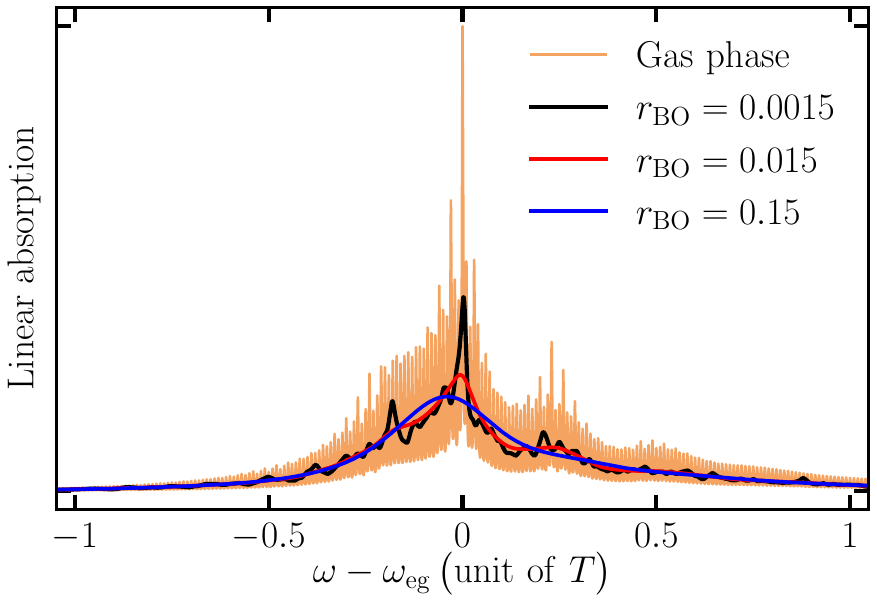}
\caption{
The evaluated absorption spectra with different $r_{\BO}$ in uncorrelated scenario. See the text for details of parameters.
 All values are uniformly scaled by the peak amplitude of gas-phase case.}
\label{fig4}
\end{figure}

Figure \ref{fig4} focuses on the solvent effect in term of BO parameter, $r_{\BO,i}\equiv \Gamma_{ii}/(2\Omega_i)$, whereas gas phase scenario ($r_{\BO,i}=0$) has an analytical solution.\cite{Yan865908}
To better visualize the vibronic progression and underlying thermal effect, we consider low--frequency vibrational modes, with $(\Omega_1, \Omega_2)/T=(0.2, 0.24)$, $(\Omega'_1,\Omega'_2)=(0.21,0.23)$ in the ground and excited states, respectively.
The intrinsic richness of the spectral progression originates from the energy level disparities between the ground and excited states induced by Duschinsky rotation.
Figure \ref{fig4} is depicted concerning varied values of $r_{\BO}$ in the uncorrelated scenario, with $\theta=\pi/3$.
Here, we set $r_{\BO}=r_{\BO,1}=r_{\BO,2}$ for both modes.
For the gas phase, $r_{\BO}=0$, the spectrum is evaluated via the analytical solution in Ref.\,\onlinecite{Yan865908}, phenomenologically broadened by an electronic dephasing, with $\w_{eg}$ being replaced by $\w_{eg}-i\gamma$, where $\gamma/T = 0.0016$. 
Figure \ref{fig4} illustrates the solvent effect on absorption spectra, 
which is not phenomenological but evaluated based on quantum dissipation theory, here via \Eq{gen_DEOM} with \Eq{alpha}. The method is applicable to realistic molecules in solvents combined with quantum chemistry calculations and molecular dynamics.\cite{Che11194508, Wan19e1375,Wan226391,Zha24e1727, Su254107}

\subsection{Electron transfer dynamics}\label{numB}

\begin{figure}[!t]
\centering
\includegraphics*
[width=
0.48\textwidth]
{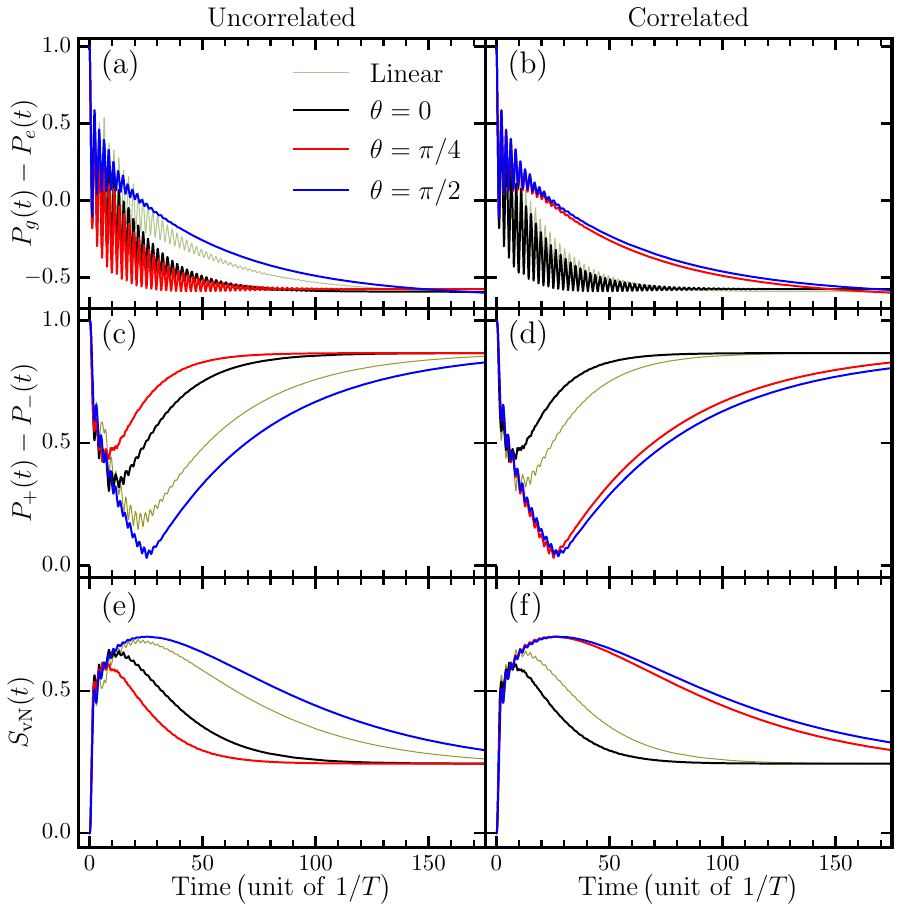}
\caption{Transfer evolution in terms of $P_{g}-P_{e} $ [panels (a) and (b)] and $P_+-P_-$ [panels (c) and (d)], together with the  evolution of von Neumann entropy [panels (e) and (f)]. See the text for details.  Shown in the left and right  panels  are the  uncorrelated and correlated scenarios, respectively.}
\label{fig5}
\end{figure}

\begin{figure*}[!t]
\centering
\includegraphics
[width=
0.75\textwidth]
{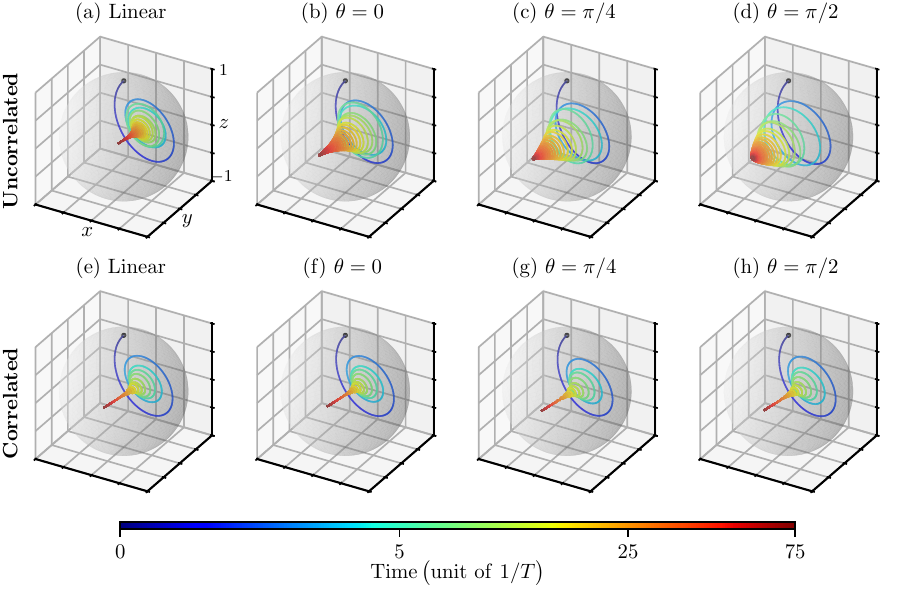}
\caption{The evolution of $\rho_{\tS}(t)$ represented via Bloch sphere with varied Duschinsky rotation angles,  as well as the linear coupling case. Shown in the upper and lower panels are the uncorrelated and correlated scenarios, respectively.}
\label{fig6}
\end{figure*}

\begin{figure*}[!t]
\centering
\includegraphics
[width=0.75
\textwidth]
{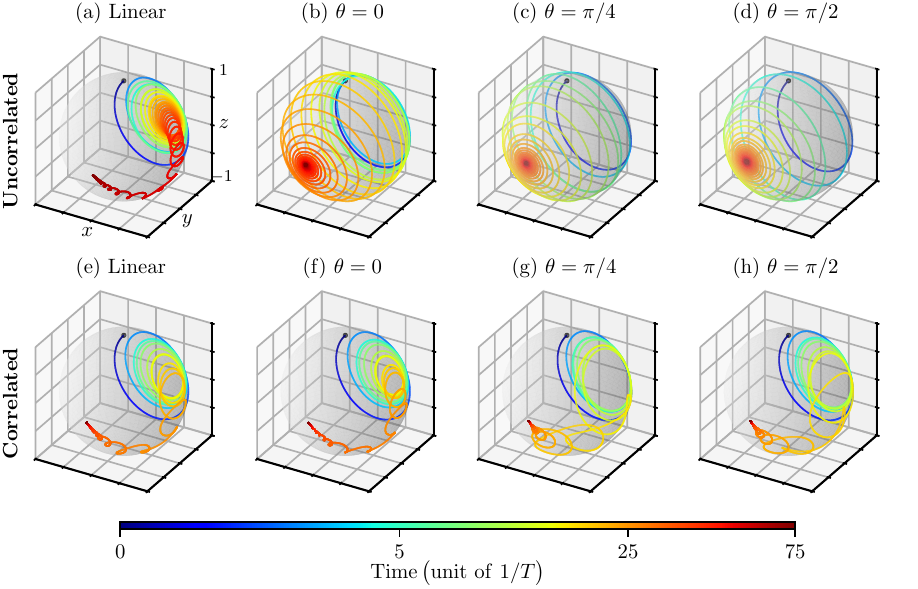}
\caption{The evolution of instantaneous diagonalized basis, $|\psi_{+}(t)\ra$, with varied Duschinsky rotation angles,  as well as the linear coupling case. Shown in the upper and lower panels are the uncorrelated and correlated scenarios, respectively.}
\label{fig7}
\end{figure*}

In this subsection, we consider the coherent ET process. We set $\w_{eg}/T=-2$ and $V/T=1$ [cf.\,\Eq{Htot}]. The other parameters are the same as those in \Fig{fig1}.
The system was initially in the ground state $|g\ra$, with $P_g(t)-P_e(t)\equiv \la g|\rho_{\tS}(t)|g\ra-\la e|\rho_{\tS}(t)|e\ra$, the population difference between the ground and excited states, initially being $1$.
The evaluated $\rho_{\tS}(t)$ can be instantaneously diagonalized into
\be \label{wy_40_0425}
\rho_{\tS}(t)\!=P_{+}(t)|\psi_+(t)\ra\la \psi_+(t)|+\!P_{-}(t)|\psi_-(t)\ra\la \psi_-(t)|,
\ee
where $|\psi_{\pm}(t)\ra$ denotes the  instantaneous canonical basis. This directly leads to
the von Neumann entropy,
\begin{align}
S_{\rm vN}(t)&\equiv -{\rm tr}_{\tS}[\rho_{\tS}(t)\ln \rho_{\tS}(t)]\nl
&=-P_{+}(t)\ln P_{+}(t)-P_{-}(t)\ln P_{-}(t).
\end{align}

Figure \ref{fig5} depicts the transfer evolution in terms of $P_g(t)-P_e(t)$ [panels (a) and (b)]  and $P_{+}(t)-P_{-}(t)$ [panels (c) and (d)] in both uncorrelated and correlated scenarios, together with the von Neumann entropy evolution [panels (e) and (f)].
In both linear (thin-olive) and $\theta=0$ (black) cases, the correlated solvent effects on BO modes accelerate the transfer process to equilibrium, as observed comparing the right panels to the left ones.
For $\theta=\pi/4$ (red), i.e.\ $\hat q'_1\sim \hat q_1-\hat q_2$ and $\hat q'_2\sim \hat q_1+\hat q_2$ [cf.\,\Eqs{coorditrans1} and (\ref{rotation_m})], the solvent-induced BO-mode correlation significantly suppresses the ET rate.
In contrast, the case of $\theta=\pi/2$ (blue), $\hat q'_1\sim -\hat q_2$ and $\hat q'_2\sim \hat q_1$, which is of the slowest ET behavior, exhibits minimal differences between correlated and uncorrelated scenarios.
As demonstrated above, the transfer dynamics exhibit the complex interplay between Duschinsky rotation and 
solvent-induced BO-mode correlation. 

The complete description of  $\rho_{\tS}(t)$ can go with Bloch sphere representation, in terms of
\be \label{wy_42_0425}
r_{i}(t)\equiv {\rm tr}_{\tS}[\hat \sigma_{i}\rho_{\tS}(t)], \quad i=x,y,z
\ee
with $\hat \sigma_{i}$ being the Pauli matrices. 
The Bloch sphere is a geometric representation of the state space of a two-level quantum system (qubit). Any pure state can be written as
\be 
|\psi\rangle_{\tS}= \cos\left(\frac{\vartheta}{2}\right)|g\rangle + e^{i\varphi}\sin\left(\frac{\vartheta}{2}\right)|e\rangle,
\ee
where \( \vartheta \in [0,\pi] \) and \( \varphi \in [0,2\pi) \), and corresponds to a point on the surface of the unit sphere in \( \mathbb{R}^3 \) with coordinates \( (r_x, r_y, r_z) = (\sin\vartheta \cos\varphi, \sin\vartheta \sin\varphi, \cos\vartheta) \). 
Two endpoints of a diameter correspond to two orthogonal wavefunctions.

More generally, any mixed state can be represented by a density matrix
\be 
\rho_{\tS} = \frac{1}{2}(\hat I + \bm{r} \cdot \bm{\hat \sigma}),
\ee
where \( \bm{r} = (r_x, r_y, r_z) \) is called the Bloch vector, \( \bm{\hat \sigma} = (\hat \sigma_x, \hat \sigma_y, \hat \sigma_z) \) are the Pauli matrices, and the condition \( |\bm{r}| \leq 1 \) ensures the positivity of \( \rho_{\tS} \). Pure states lie on the surface of the sphere (i.e.\,\( |\bm{r}| = 1 \)), while mixed states occupy the interior (\( |\bm{r}| < 1 \)), and the completely mixed state of maximum entropy  corresponds to \( \bm{r} = 0 \).
If a mixed state $\rho_{\tS}$ is a convex combination,
\begin{equation}
\rho_{\tS} = p \rho_{1} + (1-p) \rho_{2},
\end{equation}
the Bloch vector $\bm{r}$ is the weighted average:
\begin{equation}
\bm{r} = p \bm{r}_1 + (1-p) \bm{r}_2,
\end{equation}
i.e., ${\bm r}$ is located on the line segment connecting ${\bm r}_1$ and ${\bm r}_2$, satisfying $|{\bm r}-{\bm r}_2|/|{\bm r}_1-{\bm r}_2|=p$ (law of the lever).
These properties help us visualize the transfer dynamics. 
The Bloch sphere representation can be generalized to  $N$-level systems, with the Bloch vector belonging to an ($N^2-1$)--dimension real space.\cite{Kim03339}

\begin{figure*}[!t]
\centering
\includegraphics
[width=0.75
\textwidth]
{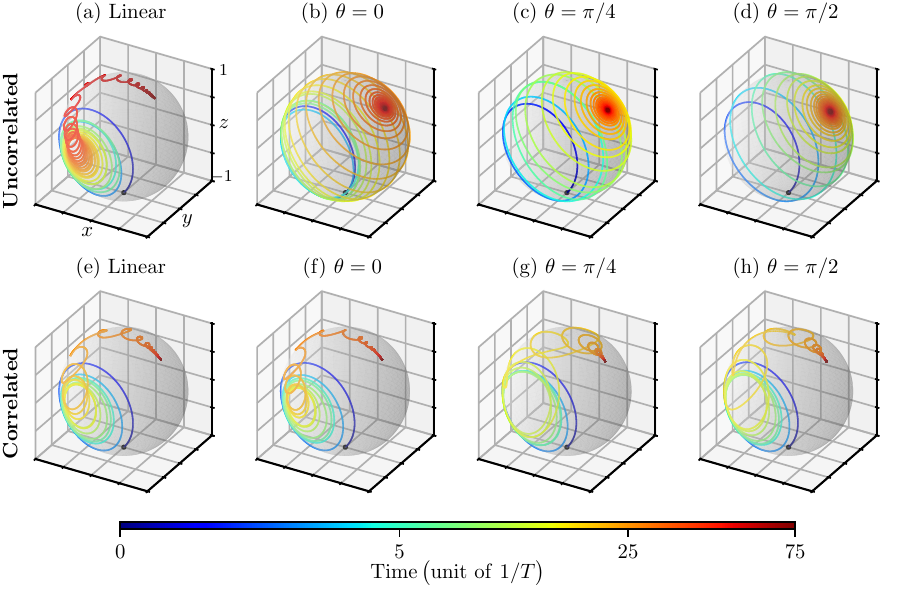}
\caption{The evolution of instantaneous diagonalized basis, $|\psi_{-}(t)\ra$, with varied Duschinsky rotation angles,  as well as the linear coupling case. Shown in the upper and lower panels are the uncorrelated and correlated scenarios, respectively.}
\label{fig8}
\end{figure*}

Figure \ref{fig6} depicts the evolution of $\rho_{\tS}(t)$, for uncorrected and correlated scenarios,    in term of \( \bm{r} = (r_x, r_y, r_z) \) by the Bloch sphere representation with varied Duschinsky rotation angles, as well as the linear coupling case. Demonstrated in \Fig{fig7} and \Fig{fig8} are the corresponding evolutions of the instantaneously diagonalized basis $|\psi_{+}\ra$ and $|\psi_{-}\ra$, respectively [cf.\,\Eq{wy_40_0425}].
These three figures can be interpreted via law of the lever, together with \Fig{fig5}(c) 
and \Fig{fig5}(d).
That is at each moment the point in \Fig{fig6} relates to the corresponding points in \Fig{fig7} and \Fig{fig8} via law of the lever, with lever lengths being $P_{\pm}(t)$ which have been indicated in \Fig{fig5}(c) 
and \Fig{fig5}(d), for uncorrected and correlated scenarios, respectively.
As illustrated, the coherent ET dynamics exhibit distinct Duschinsky--angular dependence under uncorrelated versus correlated scenarios.
For the uncorrelated scenario, 
the coherence time is enhanced with Duschinsky rotation angle $\theta$ increased.
In the correlated scenario, the decoherence is uniformly accelerated, and demonstrates angular insensitivity. These are consistent with the spectral behavior exhibited in \Fig{fig1}.
It is also observed that the Bloch sphere representation provides remarkable advantages for intuitively studying the coherent ET dynamics and furthermore entangled properties in qubits. Its extension to multi-level systems also facilitates the development of quantum computing models based on various numeral systems.\cite{Kim03339}

\section{Summary}
\label{thsec4}

We apply the extended dissipaton-equation-of-motion (ext-DEOM) method to simulate the linear absorption spectra and electron transfer (ET) dynamics involving solvent-induced BO-mode correlation and Duschinsky effects. The ext-DEOM is an exact and non-Markovian, non-perturbative approach to handle nonlinear bath couplings which are caused in this paper mainly by the Duschinsky rotation.
Elaborated also is the detail on how to disassemble
the total composite Hamiltonian, which can be constructed for realistic solvated molecules via quantum chemistry calculations, into system--plus--bath form with characterized bath coupling descriptors.
The complexity and the importance of a comprehensive interplay between solvent-induced BO-mode correlation and Duschinsky effects are illustrated. 
It shows that  precise evaluation is necessary for the reliable analysis and prediction of optical spectra and coherent transfer dynamics
in complex systems.

Finally, it is worth noting that the ext-DEOM method employed in this study can be not only applied  to spectroscopy and transfer dynamics, but also extended to address problems in both equilibrium and non-equilibrium thermodynamics.\cite{Che23074102} Not only its bosonic but also fermionic versions \cite{Xu17395,Xu18114103,Su23024113} have been developed and the latter has been utilized to study the characteristics of an adatom in functional materials.\cite{Su254107} The method can also be applied to study and modulate the effects of high-order environmental noise in quantum computing devices.\cite{Moc21186804,Han24226001,Bla21025005,Kra19021318}

\begin{acknowledgments}
Support from the Ministry of Science and Technology of China (Grant No.\ 2021YFA1200103), the National Natural Science Foundation of China (Grant Nos.\  22173088, 22373091, 224B2305), and the Innovation Program for Quantum Science and Technology (Grant No.\ 2021ZD0303301) is gratefully acknowledged.
\end{acknowledgments}

\end{document}